\documentclass{nature}

\usepackage{float}
\usepackage{graphicx,color}
\usepackage{dcolumn}
\usepackage{amsmath}
\usepackage{epsfig}
\usepackage[left]{lineno}
\usepackage{caption}

\title{Superconductivity above 30 K due to the introduction of oxygen element in CaFeAsF}

\author{Yixin Liu$^{1,2}$, Teng Wang$^{1,3}$, Zulei Xu$^{1,2}$, Da Jiang$^{1,4}$, Yi Zhao$^{3}$, Yanpeng Qi$^{3,5,6}$, Xiaoni Wang$^{1,2}$, Ming Yang$^{1,7}$,  Mao Ye$^{1,2}$,  Wei Peng$^{1,2}$, and Gang Mu$^{1,2,\ast}$}

\begin{document}

\maketitle

\begin{affiliations}
 \item State Key Laboratory of Materials for Integrated Circuits, Shanghai Institute of Microsystem and
Information Technology, Chinese Academy of Sciences, Shanghai 200050, China
 \item University of Chinese Academy of Sciences, Beijing 100049, China
   \item School of Physical Science and Technology, ShanghaiTech University, Shanghai 201210, China
 \item Institute for Frontiers and Interdisciplinary Sciences, Zhejiang University of
Technology, Hangzhou 310014, China
 \item ShanghaiTech Laboratory for Topological Physics, ShanghaiTech University, Shanghai 201210, China
 \item Shanghai Key Laboratory of High-resolution Electron Microscopy, ShanghaiTech University, Shanghai 201210, China
  \item School of Microelectronics, Shanghai University,
Shanghai 200444, China\\
$^\ast$Correspondence author: mugang@mail.sim.ac.cn
\end{affiliations}
\clearpage
\begin{abstract}
Exploring new unconventional superconductors is of great value for
both fundamental research and practical applications. It is a
long-term challenge to develop and study more hole-doped
superconductors in 1111 system of iron-based superconductors. Here
we report the discovery of superconductivity with the critical
transition temperature up to 30.7 K in the compound CaFeAsF by a
post-annealing treatment in air atmosphere. The superconducting
behaviors are verified in both the single-crystalline and
polycrystalline samples by the resistance and magnetization
measurements. The analysis by combining the depth-resolved
time-of-flight secondary ion mass spectrometry (TOF-SIMS) and X-ray
photoelectron spectroscopy (XPS) measurements show that the
introduction of oxygen elements and the consequent changing in Fe
valence by the annealing treatment may lead to the hole-type doping,
which is the origin for the occurrence of superconductivity.
Our result paves the way for further in-depth investigations on the hole-doped 1111 system in iron-based superconductors.
\end{abstract}


\clearpage
\section{Introduction}
As the second class of high-temperature superconducting (SC) system
after copper-based superconductors in terms of critical transition
temperature ($T_c$) at ambient pressure, iron-based superconductors
(IBSs)\cite{LaFeAsO} show great potential in both basic
research\cite{QMSi2008,Mazin2011,Johnston2010,ReviewHosono2015} and
practical applications\cite{HOSONO2018278}. In general, the
discovery of new superconductors in the field of unconventional
superconductivity often leads to unexpected new insights into the
physics of superconductivity. For example, the discovery of
superconductivity in A$_x$Fe$_{2-y}$Se$_2$ (A = K, Rb,
Tl)\cite{XLChen2010,Fang2011} and monolayer FeSe on
SrTiO$_3$\cite{monoFeSe} led to the realization that
superconductivity with relatively high-$T_c$ may also exist in
systems where the hole-type Fermi surface disappears
completely\cite{DLFeng2011,HDing2011,XJZhou2011} and the Fermi
surface nesting is not the key factor for occurrence of high-$T_c$
superconductivity in this system\cite{JPHu2012}. After sixteen years
of continuous efforts, a wide variety of iron-based superconducting
materials with different crystal structures and $T_c$ have been
discovered\cite{XLChen2010,Fang2011,monoFeSe,XHChen43K,RenZA2008,WANG2008538,PRB060505,FeSe,122,Qi2008,Zhu2009,12442-1}.
To date, it has become increasingly difficult to discover new
superconductors in this system by conventional means, both in terms
of new structures and new ways of inducing superconductivity
(including chemical doping, pressurization, etc.).

It's worth drawing our attention to the 1111 system ReFeAsO (Re =
rare-earth element), which is the earliest discovered
subclass\cite{LaFeAsO} and reveals the highest $T_c$ ($\approx$ 55
K)\cite{RenZA2008} in the bulk materials of the IBSs. Both
electron-\cite{LaFeAsO,XHChen43K,RenZA2008} and
hole-doping\cite{Wen_2008,B815830D,Mu2009} can induce
superconductivity by substituting F and Sr to the site of O and rare
earth elements, respectively. A noteworthy state of affairs is that
the materials and the research on the hole-doping side are
relatively rare. To develop more hole-doped superconductors in 1111
system of iron-based superconductors is a long-term challenge. The
AeFeAsF (Ae = Sr, Ca, Eu) compound is called the fluorine-based 1111
system\cite{CaFeAsF,SrFeAsF-1,SrFeAsF-2}, which reveals the same
crystal structure as ReFeAsO by replacing the ReO layer with AeF
layer. The pristine AeFeAsF is not superconducting\cite{SrFeAsF-1}.
Typically, superconductivity can be induced by chemical doping (e.g.
Ca$_{1-x}$Nd$_x$FeAsF\cite{ChengEPL2009} and
CaFe$_{1-x}$Co$_x$AsF\cite{Matsuishi2009}) or applying
pressure\cite{Pressure2010,Gao2018}. In recent years, due to the
successful acquisition of high-quality single-crystalline
samples\cite{Ma2015,Ma2016}, people have carried out in-depth and
extensive research on the fluorine-based 1111-type compound CaFeAsF
\cite{Taichi2018,Xiao2016,CaFeAsFCo-Hc2,Xu2018,Ma2018,Mu2018,Terashima2022-1,Terashima2022-2}.

In this fluorine-based 1111 system, in most cases the dominant
charge carrier is electron-type\cite{ChengEPL2009,Matsuishi2009}.
Many efforts have also been made to explore hole-doped
superconductors in this system. Research on this issue has
struggled, probably due to limitations in terms of chemical
properties. To our knowledge, the only successful case in this
regard is the doping of Na at the site of Ca atoms in the CaFeAsF
compound\cite{Shlyk2014}. We have attempted to use conventional
methods to dope O elements in the F-site, but it has been proven to
be ineffective. In order to discover new superconductors with the
hole-doped charge carriers in the fluorine-based 1111 system, here
we report a new route to introduce O atoms to the CaFeAsF compound
by the post-annealing treatment in air atmosphere. Firstly, we will
present the results obtained on single-crystalline samples,
including crystal structure, resistance, magnetization data, as well
as the behaviors of upper critical field. Secondly, in order to
further confirm the existence of superconductivity, systematical
studies were conducted on the influence of annealing temperature on
critical transition temperature and SC volume fraction in powder
polycrystalline samples. Finally, the elemental content and the Fe
valence were characterized by the depth-resolved time-of-flight
secondary ion mass spectrometry (TOF-SIMS) and X-ray photoelectron
spectroscopy (XPS) measurements. The mechanism for the occurrence of
superconductivity in the present system is discussed based on the
observations.

\section{Materials and methods}

\subsection{Sample preparation and annealing}
The CaFeAsF single crystals were grown by the self-flux method.
High-purity powders of CaAs, FeF$_2$, and Fe were used as raw
materials. Additional 4 times of CaAs were added as the flux. The
raw chemical materials are thoroughly ground and mixed in an agate
mortar, and then loaded into an alumina crucible. Then the raw
materials are sealed in a quartz tube that has been evacuated. The
single crystals were obtained in a slow cooling process from
1230$^o$C to 900$^o$C with a rate of 2$^o$C/hr. Due to the layered
structure of the present system, the crystal has a higher growth
rate in the intra-layer direction. Therefore, the obtained single
crystals naturally present a shape of flake and exhibit a $c$-axis
orientation (see Fig. 1(b)). The detailed growth process has been
reported in the previous work\cite{Ma2015}.

The polycrystalline samples were prepared using the solid-state
reaction. The raw reagents, which are the same as that used in the
growth except for the flux, were thoroughly ground and mixed in an
agate mortar. Three sintering temperatures, 920$^o$C, 980$^o$C, and
1010$^o$C, were used in our experiments. By checking the quality and
purity (see Fig. S1), samples obtained at 1010$^o$C were used for
the following annealing treatment.

The single-crystalline samples were annealed in air at 330 $^o$C for
18 hrs. For the polycrystalline samples, the annealing temperature
ranges from 300 to 450 $^o$C. Two different annealing time, 12 and
18 hrs, were adopted. In order to identify the decisive gas
components in the annealing process, we also carried out experiments
in different atmospheres. Besides the air atmosphere mentioned
above, we checked the annealing effect of CaFeAsF single crystals in
high purity oxygen (O$_2$), the mixture of nitrogen and oxygen that
mimics the proportions of air (N$_2$+O$_2$), and carbon dioxide gas
(CO$_2$).

\subsection{Characterization}
The crystal structure of the CaFeAsF samples, including
single-crystalline and polycrystalline ones, were examined by a
DX-2700-type X-ray diffractometer using Cu $K_{\alpha}$ radiation.
The element contents were measure using the time-of-flight secondary
ion mass spectrometry (ION-TOF GmbH, TOF-SIMS 5-100). The valence of
Fe elements was evaluated using the X-ray photoelectron spectroscopy
(Thermo Fisher Scientific, Escalab 250Xi). To obtain depth
dependence of information, the single-crystalline samples were
etched using Cs and Ar ions when carrying out TOF-SIMS and XPS
measurements respectively. In the TOF-SIMS measurements, the etch
area is 180 $\times$180 $\mu$m$^2$. The energy of the ion beam is 1
keV. The microstructure of the single-crystalline samples was
characterized by the Cs-corrected transmission electron microscopy
(JEOL, JEM-ARM300F).

\subsection{Physical property measurements}
The electrical resistance were measured on the physical property
measurement system (Quantum Design, PPMS) by a standard four-probe
method. The silver paste was used to fabricate the electrodes. The
external magnetic field of up to 9 T was applied in two orientations
($H\parallel c$ and $H\parallel ab$) and perpendicular to the
electric current. The magnetization measurements were carried out on
the magnetic property measurement system (Quantum Design, MPMS 3).
For the magnetization measurements on single-crystalline samples,
the magnetic field was applied parallel to the $ab$-plane to
minimize the effect of the demagnetization effect. The specific heat
were measured using a thermal relaxation technique on the physical
property measurement system (Quantum Design, PPMS).

\section{Results and Discussion}
\subsection{Superconductivity in air-annealed single-crystalline
CaFeAsF}

\begin{figure}\centering
\includegraphics[width=12cm]{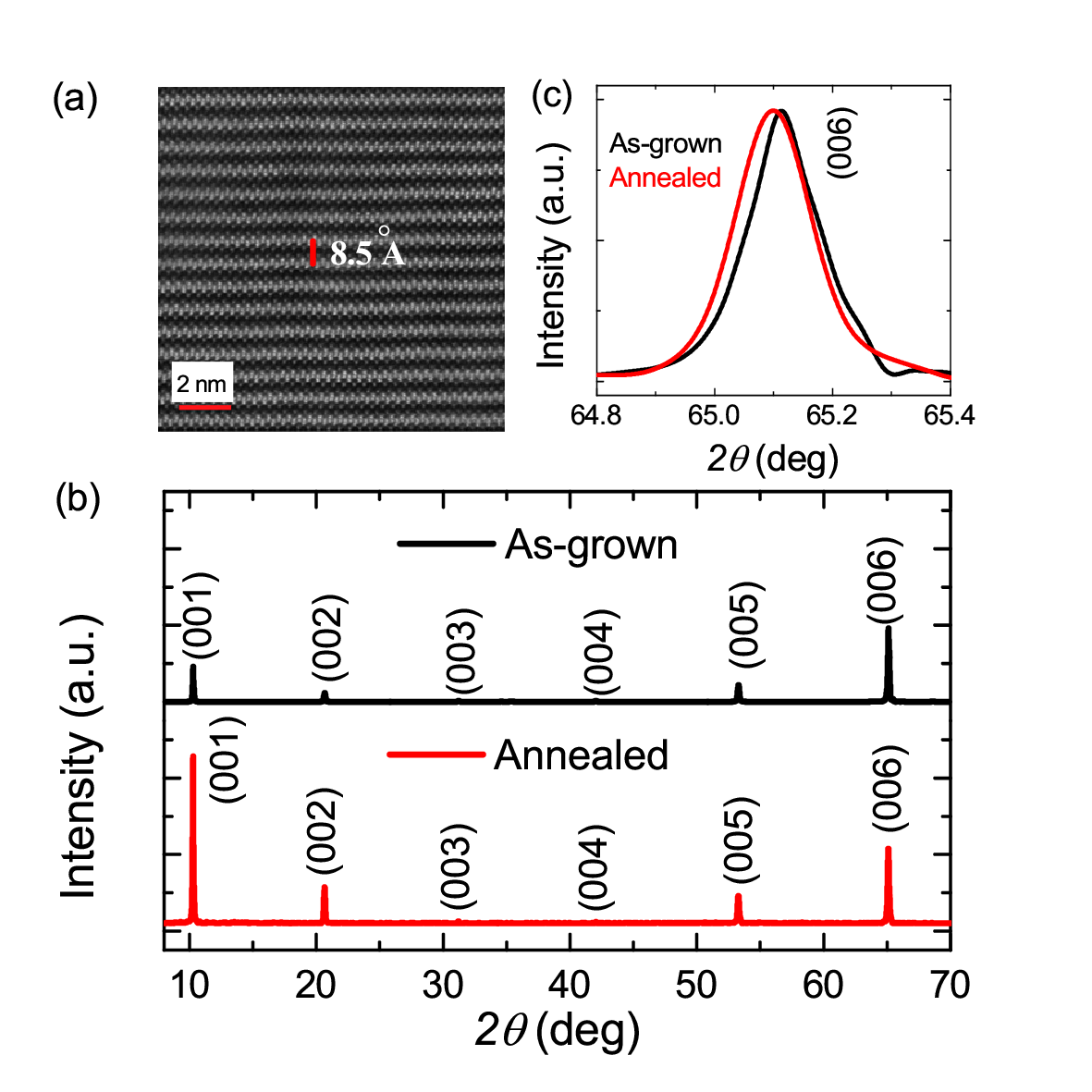}
\caption {(a) Cs-corrected TEM image of an as-grown CaFeAsF single
crystal. The layer spacing is indicated by the vertical red line.
(b) XRD patterns of the as-grown and annealed single-crystalline
CaFeAsF. (c) An enlarged view of the XRD data near (006) peak.}
\label{fig1}
\end{figure}

As shown in Fig. 1(a), the transmission electron microscopy (TEM)
image of CaFeAsF single crystal exhibits the atomic arrangement with
a clear layered feature. The weak bending feature on the layered
structure is due to the slight movement of the sample during the
process of data acquisition. The layer spacing is about 8.5 $\AA$,
which is consistent with that determined from the previous
report\cite{Ma2015}. The X-ray diffraction (XRD) patterns show that
both annealed and as-grown samples exhibit good crystallinity. As
shown in Fig. 1(b), only sharp peaks with the index (00$l$) can be
observed, suggesting a high $c$-axis orientation. The position of
the diffraction peaks of the annealed sample show a very slight
shift ($\sim$ 0.016$^o$) to the left as compared to the as-grown
sample (see Fig. 1(c)), suggesting a slight expansion of the crystal
lattice due to the annealing treatment.

\begin{figure}\centering
\includegraphics[width=14cm]{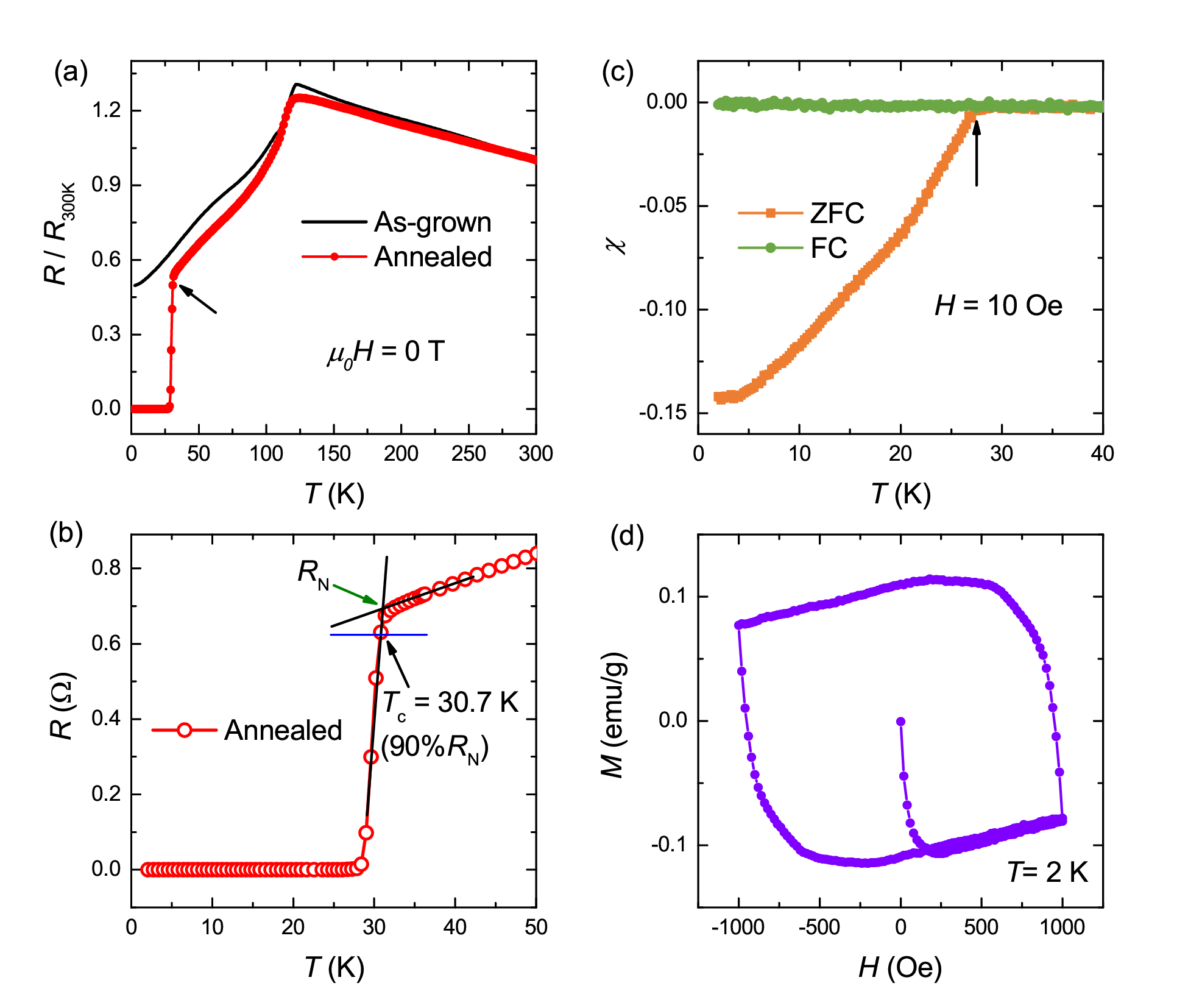}
\caption {Superconductivity in single-crystalline CaFeAsF annealed
in air atmosphere. (a) Temperature dependence of normalized
resistance of the as-grown (black) and annealed (red) CaFeAsF single
crystals under zero magnetic field. (b) Resistance data of the
annealed samples in the low temperature region. The arrowed lines
indicated the criterion in determining the critical transition
temperature $T_c$. (c) Magnetic susceptibility of the annealed
sample measured in zero-field-cooled (ZFC) and field cooled (FC)
modes. (d) The isothermal hysteresis loop of magnetization for the
annealed sample at 2 K. The black arrows in (a) and (c) indicate the
onset transition point in the $R/R_{300K}$-$T$ and $\chi$-$T$ curves
respectively.} \label{fig2}
\end{figure}

We measured resistance data of the CaFeAsF single crystals annealed
at 330 $^o$C for 18 hrs. As shown in Fig. 2(a), the data undergoes a
clear and sharp superconducting transition in the low-temperature
region. Meanwhile, the $R/R_{300K}$-$T$ curve roughly follow a
similar trace with that of the as-grown CaFeAsF in the
high-temperature region, which reveals a structure transition at
around 121 K. Such a behavior indicates the coexistence of
superconductivity and the pristine CaFeAsF phase in the annealed
sample. In order to have a clearer view of the superconducting
transition, we show the resistance in the low temperature region in
Fig. 2(b). Using a criterion of 90\% $R_n$, where $R_n$ is the
normal-state resistance before SC transition, the onset transition
temperature $T_c$ is determined to be 30.7 K. We note that such a
value is about 10 K higher than that observed in Co-doped
CaFeAsF\cite{Matsuishi2009,Ma2016}.

The occurrence of superconductivity is further confirmed by the
magnetic susceptibility ($\chi$) measurements, see Fig. 2(c). The
magnetic field was applied parallel to the $ab$-plane of the crystal
to minimize the effect of the demagnetization effect. The onset
transition point in the $\chi$-$T$ curve is 27.5 K, which is very
close to the zero-resistance temperature. The value of magnetic
susceptibility actually reflects the SC volume fraction of the
sample, because the magnetic susceptibility of a fully SC sample
should be -1 based on the Meissner effect. Thus, the SC volume
fraction of this samples is estimated to be 14.2\% from the value of
$\chi$ at 2 K. This is consistent with the inference from the
normal-state resistance data (see Fig. 2(a)) that the SC state
coexists with the pristine phase of CaFeAsF. The $M$-$H$ curve
measured at 2 K shows a typical hysteresis loop of the type-II
superconductors (see Fig. 2(d)).

\begin{figure}\centering
\includegraphics[width=12cm]{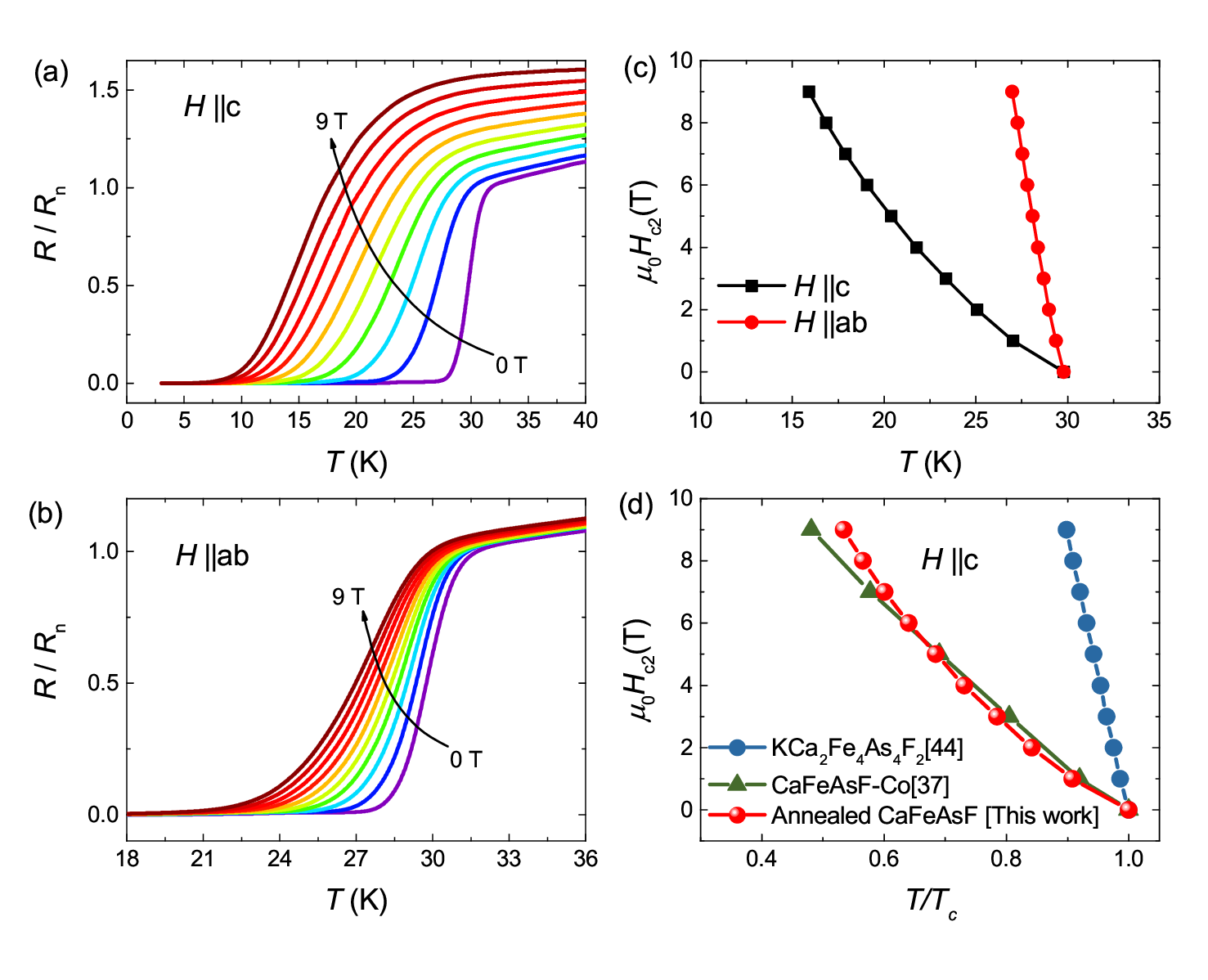}
\caption {Upper critical field of annealed CaFeAsF single crystals.
(a, b) The normalized resistance as a function of temperature under
the magnetic field up to 9 T with $H||c$ and $H||ab$, respectively.
The interval of the magnetic field is 1 T. (c) Upper critical fields
$\mu_0 H_{c2}$ as a function of temperature for two field
orientations. (d) Comparison of $\mu_0 H_{c2}$ as a function of
$T/T_c$ among the three systems
KCa$_2$Fe$_4$As$_4$F$_2$\cite{12442}, Co-doped
CaFeAsF\cite{CaFeAsFCo-Hc2}, and the air-annealed CaFeAsF studied in
the present work.} \label{fig3}
\end{figure}

To study the upper critical field and its anisotropy, we performed
the measurements on the temperature dependence of electrical
transport with the magnetic field along two different orientations:
$H\parallel c$ and $H\parallel ab$. As shown in Figs. 3(a) and (b),
for both field orientations, the SC transition point shifts to lower
temperature with the increase of the magnetic field. It is notable
that such a shift is much more quickly when the magnetic field is
applied along the $c$ axis, revealing a significant anisotropic
characteristic. Using the criteria of 50\%$\rho_n$, the values of
upper critical $\mu_0 H_{c2}$ are extracted, see Fig. 3(c). The
$\mu_0H_{c2}$-$T$ data reveals a steeper slope near $T_c$ with $H
\parallel ab$ than that with $H \parallel c$, further verifies the
anisotropic feature of the upper critical field. Quantitatively, the
slope of the tangent in the $H_{c2}$-$T$ curves near $T_c$,
$d\mu_0H_{c2}/dT|_{T_c}$, is -2.53 and -0.36 T/K for the
orientations of $H\parallel ab$ and $H\parallel c$ respectively,
which gives rise to an anisotropy parameter of $\Gamma$ = 7. It is
worth noting that the out-of-plane upper critical field shows an
upward trend with cooling, which is inconsistent with the nearly
linear behavior of $H_{c2}$ relative to $T$ near $T_c$ in
conventional superconductors\cite{Werthamer1966,WANG20231354223}. In
Fig. 3(d), we compare this behavior with two other iron-based
superconducting materials\cite{12442,CaFeAsFCo-Hc2} using the same
criteria of 50\%$\rho_n$. The data of 12442 basically shows linear
behavior, while the Co-doped CaFeAsF and air-annealed CaFeAsF
exhibit a similar upward trend. This may reflect the unique
electronic structure of the 1111 system compared to other systems in
IBSs.

\subsection{Superconductivity in air-annealed polycrystalline
CaFeAsF} In general, polycrystalline materials have smaller grains
and larger specific surface areas than single-crystalline ones.
Therefore, in order to improve the efficiency of annealing treatment
and to increase the SC volume fraction, we also carried out
annealing experiments on polycrystalline CaFeAsF in air atmosphere.
The magnetic properties of the untreated sample and the powder XRD
patterns of annealed samples can be seen in Figs. S1 and S2. An
enlarged view of the XRD data near (102) peak is shown in the inset
of Fig. 4(a). Compared with the untreated pristine sample, the
positions of diffraction peak of the samples annealed at 350$^o$C
(see the red curve) reveals a left shift. It is worth noting that
the peak shift caused by annealing of polycrystalline samples is
0.070$^o$, which is significantly greater than that of
single-crystalline samples. This indicates that the annealing effect
in polycrystalline samples is stronger than that in
single-crystalline samples.

In Fig. 4(a), we show the temperature dependence of $\chi$ in the
ZFC mode. The data reveals a systematically evolution with annealing
temperature $T_{ann}$ and time,  When $T_{ann}$ = 300 $^o$C,
superconductivity begins to emerge, although with a rather low $T_c$
and small SC volume fraction. The maximum value of $T_c$ at around
27.7-28.5 K can be achieved when 350 $^o$C $\leq T_{ann} \leq$ 410
$^o$C, which is consistent with the results in annealed
single-crystalline samples (see Fig. 2(c)). The superconductivity is
also confirmed by the $M$-$H$ curve, as can be seen in Fig. 4(b).

In Figs. 4(c) and (d), we summarized the values of $T_c$ and SC
volume fraction as a function of the annealing temperature $T_{ann}$
derived from the data in Fig. 4(a). It can be seen that, for the
highest superconducting transition temperature, there is a wide
temperature window in the annealing conditions; while the SC volume
fraction shows a peak shape versus the annealing temperature.
Combining these two factors, the optimal annealing condition for the
polycrystalline samples is 365-380 $^o$C and 12 hrs. It is worth
noting that the highest superconducting volume fraction in the
annealed polycrystalline samples is 20.5\%, which is significantly
higher than the value obtained in the single-crystalline samples
($\sim$ 14.2\%). This confirms our previous speculation that the
structural characteristics of polycrystalline materials allow them
to exhibit more pronounced annealing effects.

\begin{figure}
\centering
\includegraphics[width=14cm]{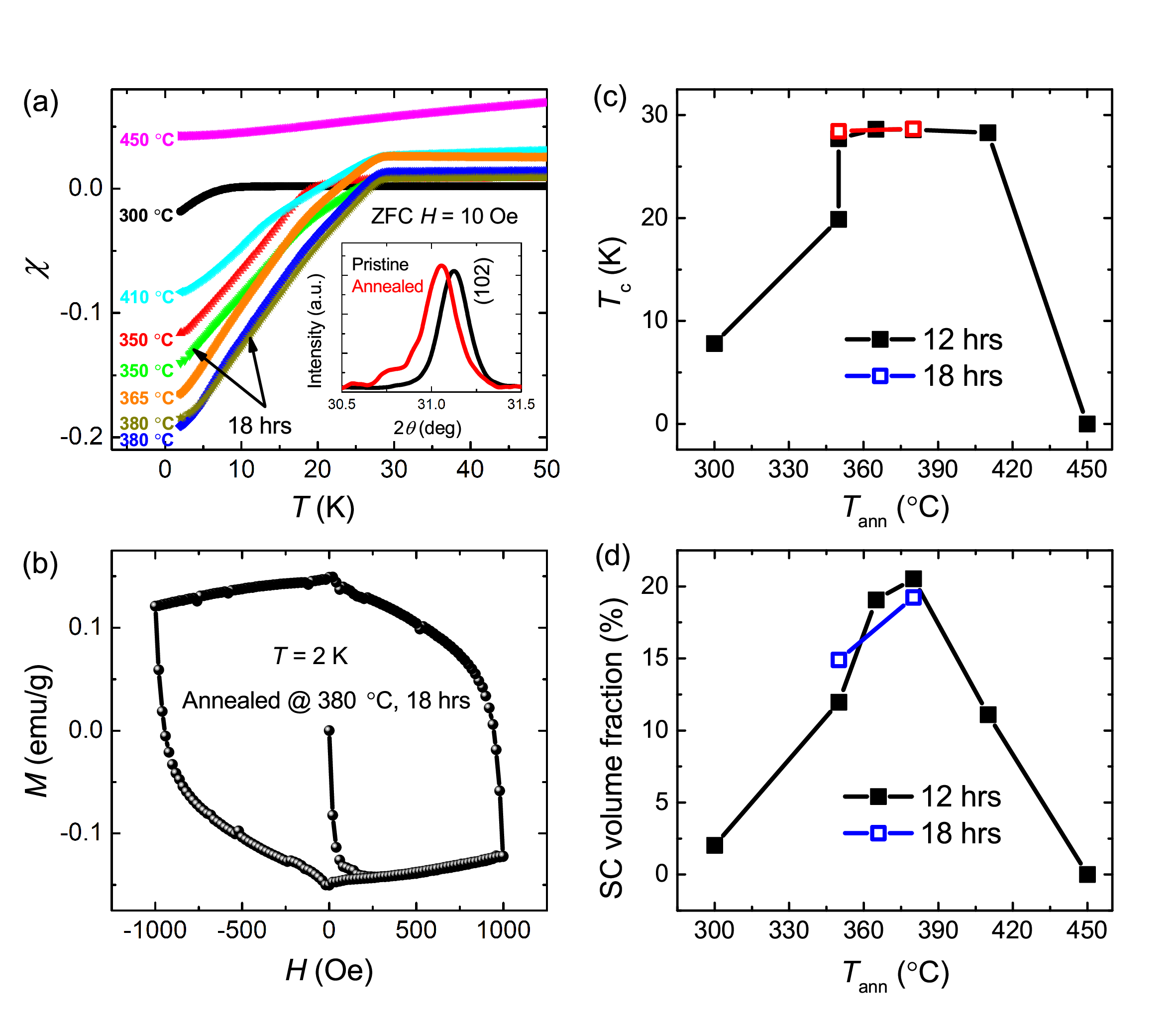}
\caption {Superconductivity in polycrystalline CaFeAsF samples
annealed under various conditions in air atmosphere. (a) Magnetic
susceptibility of the annealed samples measured with ZFC mode. The
annealing time is 12 hrs, except for the two curves indicated by the
arrows (18 hrs). Inset: an enlarged view of the XRD data near (102)
peak. (b) The isothermal hysteresis loop of magnetization for the
sample annealed at 380 $^o$C for 18 hrs. (c, d) Superconducting
critical transition temperature and the SC volume fraction as a
function of the annealing temperature.} \label{fig4}
\end{figure}

\subsection{The mechanism for the occurrence of superconductivity} It is obviously necessary to reveal the
internal mechanism for the occurrence of superconductivity in the
air-annealed CaFeAsF system. We designed comparative experiments to
identify the gas components that play a decisive role in the
annealing process. The magnetization data of samples annealed in
different atmospheres is shown in Fig. S3. Here we summarize the
main result of the annealing effect in different atmospheres in
Table 1. It can be seen that the mixture of N$_2$ and O$_2$ has a
similar effect with air and the annealing in CO$_2$ could not induce
superconductivity. Moreover, annealing in purity O$_2$ can also lead
to the occurrence of superconductivity, although with a relatively
lower $T_c$ and SC volume fraction. The SC performance of samples
annealed in pure O$_2$ is inferior to those annealed in air,
indicating that the excessively oxidizing environment caused
additional damage to the samples. Meanwhile, elemental analysis
combined with TOF-SIMS and XPS measurements showed that there was no
nitrogen elements present in the annealed samples. Combining these
information, we preliminarily deduce that the introduction of oxygen
element during the annealing process may be a key factor in the
generation of superconductivity.

\begin{table}
\centering \caption{Summary of the annealing effect on CaFeAsF
single crystals in different atmospheres. All the samples were
annealed under the same temperature and time (330 $^o$C for 18
hrs).}
\medskip
\begin{tabular}{p{4.5cm}<{\centering}p{1.5cm}<{\centering}p{1.5cm}<{\centering}p{1.5cm}<{\centering}p{1.8cm}<{\centering}}
\hline
Atmosphere & Air & O$_2$  & N$_2$+O$_2$ & CO$_2$   \\
\hline
$T_c$ $^a$  & 27.5  & 11.2  & 24.5  & Non-SC$^b$  \\
SC volume fraction  & 14.2\%  & 7.7\%  & 14.1\%  & Non-SC  \\
\hline
\end{tabular}
\\
\footnotesize{$^a$ Here the values of $T_c$ are determined from the
$\chi$-$T$ data. $^b$ Non-SC is an abbreviation of
non-superconducting.}
\end{table}

To further corroborate our inferences, we examined the
depth-resolved element content on both as-grown and annealed samples
by means of TOF-SIMS. For the as-grown sample, the oxygen content
due to the slight surface adsorption or diffusion had dropped to 2\%
of the total amount with an etch time of 120 s (see Fig. 5(a)). In
stark contrast, the annealed sample needed to be etched for more
than 1300 seconds before the oxygen content dropped to the same
level (see Fig. 5(b)). This result gives the direct evidence that
oxygen is introduced into samples during the annealing process.
Moreover, the evolution of fluorine content with depth generally
shows the opposite trend to oxygen. This indicates that the oxygen
element is mainly in competition with fluorine. As oxygen atoms
enter, fluorine atoms that originally existed in the sample
precipitate out in the form of CaF$_2$ impurities, see Fig. S2. The
cross-section images for the distribution of these two elements in
as-grown and annealed samples are shown in Figs. 5(c-f)
respectively, which give more intuitive representations of this
trend. In Figs. S5(a) and (b), we also display the three-dimensional
images of the the TOF-SIMS intensity of oxygen of the as-grown and
annealed samples respectively. From the information of etch time, we
can roughly estimate that the oxygen content affects the range of
about 150 nm from the sample surface.

\begin{figure}
\centering
\includegraphics[width=15cm]{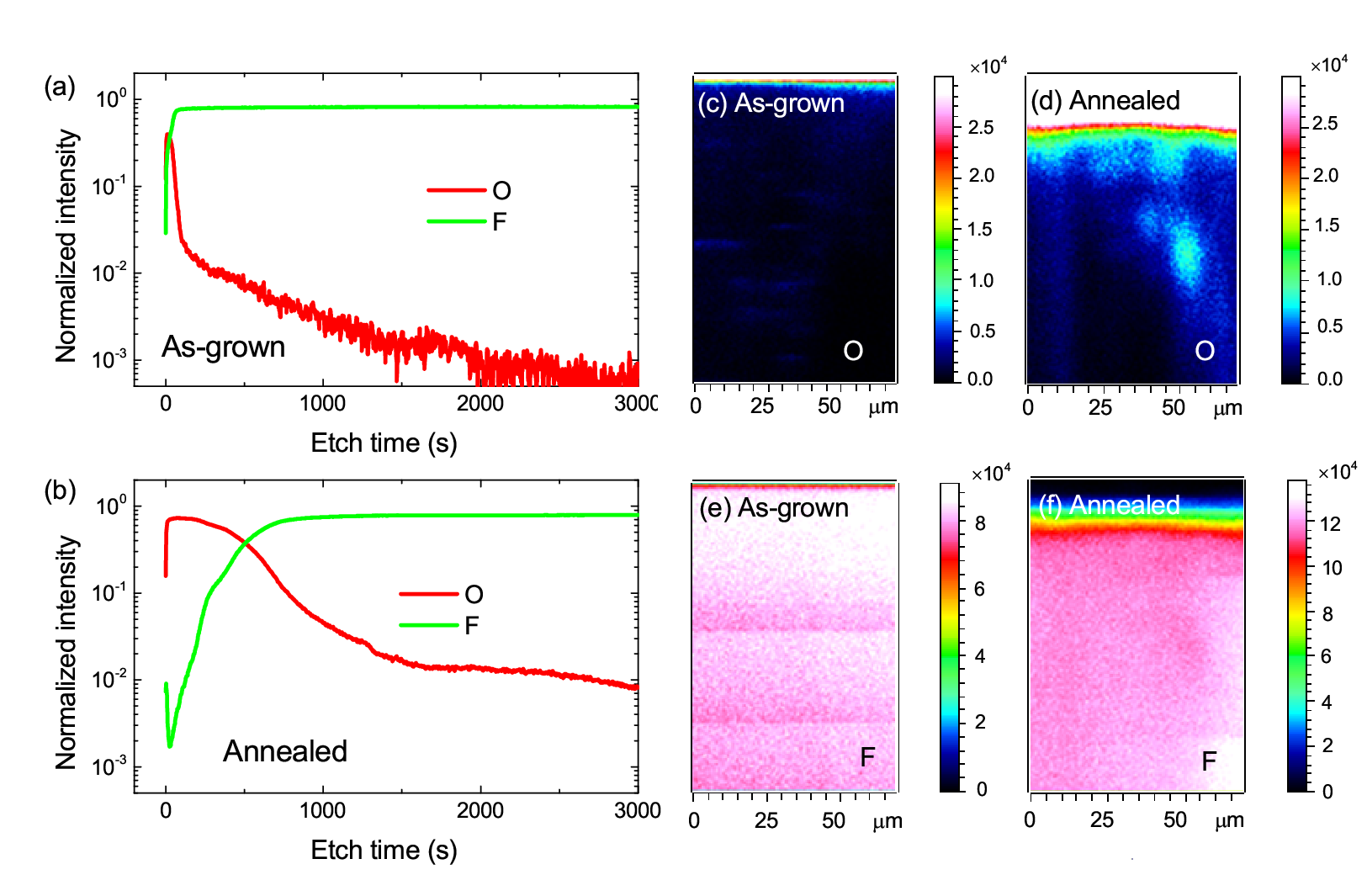}
\caption {Depth analysis of elemental content and energy spectrum
characterization in the single-crystalline samples. (a, b)
Normalized intensity of O and F elements as a function of the etch
time for the as-grown and annealed samples from the TOF-SIMS
measurements, respectively. (c, d) The cross-section images of the
TOF-SIMS intensity (unit: counts) of oxygen element for the as-grown
and annealed samples respectively. (e, f) The cross-section images
of the TOF-SIMS intensity (unit: counts) of fluorine element for the
as-grown and annealed samples respectively.} \label{fig5}
\end{figure}

We note that the observation of partial replacement of fluorine
element by oxygen is consistent with the expansion of crystal
lattice as revealed by the XRD data (see Fig. 1(c) and the inset of
Fig. 4(a)), since the radius of O$^{2-}$ is larger than that of
F$^-$. In the oxygen-based 1111 system, the partial substitution of
oxygen by fluorine element, which is in the opposite direction as
the present work, leads to the shrinkage of lattice\cite{LaFeAsO}.
Moreover, because of the limited depth of oxygen atoms entering the
sample, a relatively low SC volume fraction is rather reasonable. At
the same time, due to the gradient distribution of oxygen elements
in the depth direction as shown in Figs. 5(b) and (d), the
superconductivity in the sample is inevitably in a non-uniform
state. This explains the experimental fact that the $\chi$-$T$
curves in Fig. 2(c) and Fig. 4(a) remain declining even at low
temperatures below 5 K. Moreover, due to the rather low SC volume
fraction and the inhomogeneous feature of the sample, it is
difficult to observe SC signs in low-temperature specific heat
measurements, see Fig. S6.

One possible question is whether oxygen atoms are located in the
lattice of the material or in the interstitial sites. To test this
issue experimentally, we further annealed the samples in a vacuum
environment that had been air-annealed. If the oxygen atoms
introduced by air annealing are located in the interstitial sites,
they should be easily carried away in a vacuum environment. As can
be seen in Fig. S7, superconductivity can still be observed in the
sample annealed in vacuum at 380 $^o$C for 24 hrs. In addition, its
superconducting transition temperature and diamagnetic signal
strength are basically consistent with those of the sample without
vacuum treatment. The only observable change is that the absolute
value of the diamagnetic signal of the FC mode in the vacuum
annealed sample has decreased, indicating the presence of relatively
stronger magnetic flux pinning. This may originate from the change
in defect state in the material caused by additional vacuum
annealing. Based on these experimental results, we argue that the
oxygen atoms introduced by air annealing have entered the lattice of
the material.

\begin{figure}\centering
\includegraphics[width=12cm]{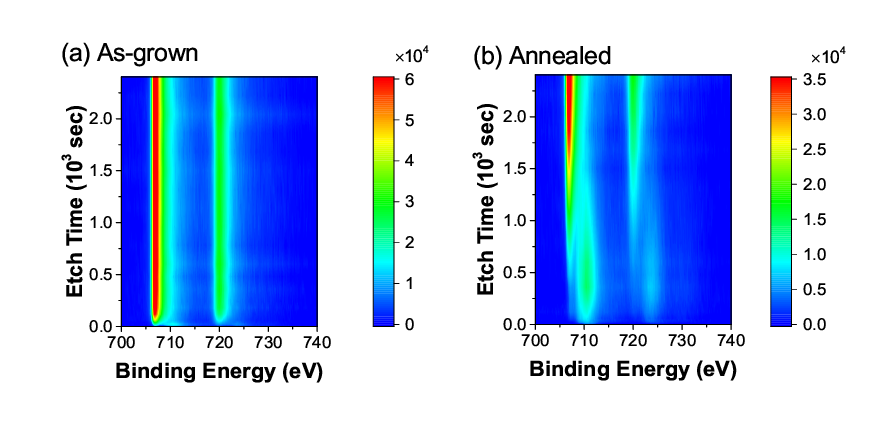}
\caption {Colormaps of the XPS spectra (unit: counts/s) of Fe 2p
state for the as-grown (a) and annealed (b) samples respectively in
a wide range of etch time.} \label{fig6}
\end{figure}

Finally, the valence of Fe ions was investigated using X-ray
photoelectron spectroscopy (XPS). The samples were etched using
argon ion with the energy of 1 keV. As shown in Fig. 6(a), there are
two main peaks located at 706.9 and 720.0 eV in the XPS spectrum of
the as-grown sample, reflecting the Fe 2p3 and 2p1 states
respectively. For the annealed sample, in addition to the
above-mentioned two peaks, two additional peaks at slightly higher
energy positions appeared on the XPS spectrum at 710.5 and 723.5 eV
when etch time of the ion beam was below about 1000 seconds, see
Fig. 6(b). It is worth noting that, due to the unavoidable reduction
during the ion etching process, it is not yet possible to derive the
precise value of Fe valence from the peak position we have obtained.
Nevertheless, the qualitative trend is certain, that is, the
annealing treatment raises the valence of Fe element. Based on this
experimental fact, we can further infer that the introduction of
oxygen elements during the annealing process probably brought about
hole-doping to CaFeAsF. And this, in turn, is the reason for the
appearance of superconductivity in this system. In this sense, in
addition to the few reported materials
(La$_{1-x}$Sr$_x$FeAsO\cite{Wen_2008} with $T_c \sim$ 25 K,
Ca$_{1-x}$Na$_x$FeAsF\cite{Shlyk2014} with $T_c \sim$ 34.5 K), we
have developed a new member of hole-doped superconductors in the
1111 system with a relatively high transition temperature ($\sim$
30.7 K). Of course, the presence of hole-type doping needs to be
confirmed by Hall effect measurements after pure superconducting
sample (without the pristine phase) is obtained in the future.

\section{Conclusions}
In summary, we discovered superconductivity with $T_c$ up to 30.7 K
in air-annealed CaFeAsF. The maximum superconducting volume fraction
is 20.5\% in the polycrystalline samples annealed at 380 $^o$C for
12 hrs. The single-crystalline sample revealed an anisotropy of
$\Gamma$ = 7 for the upper critical field. Comparative experiments
of annealing treatments under various atmospheres and
characterization analysis combining TOF-SIMS and XPS indicate that
the introduction of oxygen during annealing and the concomitant
elevation of Fe valence are the key reasons for the generation of
superconductivity. Our findings are likely to developed a new type
of hole-doped superconductivity in the 1111 system, whose $T_c$ has
exceeded that reported previously.

\section*{Conflict of interest}
The authors declare that they have no conflict of interest.

\section*{Acknowledgments}
We thank the helpful discussions with Professor G. H. Cao.
This work is supported by the National Natural Science Foundation of China (Nos. 11204338 and 52272265), National Key R\&D Program of China (No. 2018YFA0704300), and the Youth Innovation Promotion Association of the Chinese
Academy of Sciences (No. 2015187). The experimental
measurements were supported by the Superconducting Electronics
Facility (SELF) of Shanghai Institute of Microsystem and Information
Technology.

\section*{Author contributions}
G.M. designed the experiments. Y.L. and T.W. prepared the single-crystalline and polycrystalline samples, and performed the electrical transport
measurements. Y.L. performed the magnetization and TEM measurements. T.W. performed the TOF-SIMS measurements. Y.Z. and Y.Q. carried out the XPS measurements. All the authors discussed the results. G.M. analyzed
the data and wrote the paper.

\section*{Supplementary materials}
Supplementary materials to this article can be found online at
https://doi.org/xxx.

\section*{References}
\bibliographystyle{naturemag}
\bibliography{CaFeAsF-SC}

\begin{thebibliography}{10}
\expandafter\ifx\csname url\endcsname\relax
  \def\url#1{\texttt{#1}}\fi
\expandafter\ifx\csname urlprefix\endcsname\relax\def\urlprefix{URL }\fi
\providecommand{\bibinfo}[2]{#2}
\providecommand{\eprint}[2][]{\url{#2}}

\bibitem{LaFeAsO}
\bibinfo{author}{Kamihara, Y.}, \bibinfo{author}{Watanabe, T.},
  \bibinfo{author}{Hirano, M.} \& \bibinfo{author}{Hosono, H.}
\newblock \bibinfo{title}{Iron-based layered superconductor
  \textrm{La[O$_{1-x}$F$_x$]FeAs} (x = 0.05-0.12) with \textrm{$T_c$ = 26 K}}.
\newblock \emph{\bibinfo{journal}{J. Am. Chem. Soc.}}
  \textbf{\bibinfo{volume}{130}}, \bibinfo{pages}{3296--3297}
  (\bibinfo{year}{2008}).

\bibitem{QMSi2008}
\bibinfo{author}{Si, Q.} \& \bibinfo{author}{Abrahams, E.}
\newblock \bibinfo{title}{Strong correlations and magnetic frustration in the
  high \textrm{${T}_{c}$} iron pnictides}.
\newblock \emph{\bibinfo{journal}{Phys. Rev. Lett.}}
  \textbf{\bibinfo{volume}{101}}, \bibinfo{pages}{076401}
  (\bibinfo{year}{2008}).

\bibitem{Mazin2011}
\bibinfo{author}{Hirschfeld, P.~J.}, \bibinfo{author}{Korshunov, M.~M.} \&
  \bibinfo{author}{Mazin, I.~I.}
\newblock \bibinfo{title}{Gap symmetry and structure of \textrm{Fe}-based
  superconductors}.
\newblock \emph{\bibinfo{journal}{Rep. Prog. Phys.}}
  \textbf{\bibinfo{volume}{74}}, \bibinfo{pages}{124508}
  (\bibinfo{year}{2011}).

\bibitem{Johnston2010}
\bibinfo{author}{Johnston, D.~C.}
\newblock \bibinfo{title}{The puzzle of high temperature superconductivity in
  layered iron pnictides and chalcogenides}.
\newblock \emph{\bibinfo{journal}{Adv. Phys.}} \textbf{\bibinfo{volume}{59}},
  \bibinfo{pages}{803--1061} (\bibinfo{year}{2010}).

\bibitem{ReviewHosono2015}
\bibinfo{author}{Hosono, H.} \& \bibinfo{author}{Kuroki, K.}
\newblock \bibinfo{title}{Iron-based superconductors: Current status of
  materials and pairing mechanism}.
\newblock \emph{\bibinfo{journal}{Phyica C}} \textbf{\bibinfo{volume}{514}},
  \bibinfo{pages}{399--422} (\bibinfo{year}{2015}).

\bibitem{HOSONO2018278}
\bibinfo{author}{Hosono, H.}, \bibinfo{author}{Yamamoto, A.},
  \bibinfo{author}{Hiramatsu, H.} \& \bibinfo{author}{Ma, Y.}
\newblock \bibinfo{title}{Recent advances in iron-based superconductors toward
  applications}.
\newblock \emph{\bibinfo{journal}{Mater. Today}} \textbf{\bibinfo{volume}{21}},
  \bibinfo{pages}{278--302} (\bibinfo{year}{2018}).

\bibitem{XLChen2010}
\bibinfo{author}{Guo, J.} \emph{et~al.}
\newblock \bibinfo{title}{Superconductivity in the iron selenide
  \textrm{${\text{K}}_{x}{\text{Fe}}_{2}{\text{Se}}_{2}$
  $(0\ensuremath{\le}x\ensuremath{\le}1.0)$}}.
\newblock \emph{\bibinfo{journal}{Phys. Rev. B}} \textbf{\bibinfo{volume}{82}},
  \bibinfo{pages}{180520} (\bibinfo{year}{2010}).

\bibitem{Fang2011}
\bibinfo{author}{Fang, M.-H.} \emph{et~al.}
\newblock \bibinfo{title}{Fe-based superconductivity with {$T_c$ =31 K}
  bordering an antiferromagnetic insulator in {(Tl,K)Fe$_x$Se$_2$}}.
\newblock \emph{\bibinfo{journal}{Europhys. Lett.}}
  \textbf{\bibinfo{volume}{94}}, \bibinfo{pages}{27009} (\bibinfo{year}{2011}).

\bibitem{monoFeSe}
\bibinfo{author}{Wang, Q.-Y.} \emph{et~al.}
\newblock \bibinfo{title}{Interface-induced high-temperature superconductivity
  in single unit-cell {FeSe} films on {SrTiO$_3$}}.
\newblock \emph{\bibinfo{journal}{Chin. Phys. Lett.}}
  \textbf{\bibinfo{volume}{29}}, \bibinfo{pages}{037402}
  (\bibinfo{year}{2012}).

\bibitem{DLFeng2011}
\bibinfo{author}{Zhang, Y.} \emph{et~al.}
\newblock \bibinfo{title}{Nodeless superconducting gap in
  \textrm{A$_x$Fe$_2$Se$_2$ (A = K, Cs)} revealed by angle-resolved
  photoemission spectroscopy}.
\newblock \emph{\bibinfo{journal}{Nat. Mater.}} \textbf{\bibinfo{volume}{10}},
  \bibinfo{pages}{273} (\bibinfo{year}{2011}).

\bibitem{HDing2011}
\bibinfo{author}{Wang, X.-P.} \emph{et~al.}
\newblock \bibinfo{title}{Strong nodeless pairing on separate electron fermi
  surface sheets in \textrm{(Tl, K)Fe$_{1.78}$Se$_2$} probed by {ARPES}}.
\newblock \emph{\bibinfo{journal}{Europhys. Lett.}}
  \textbf{\bibinfo{volume}{93}}, \bibinfo{pages}{57001} (\bibinfo{year}{2011}).

\bibitem{XJZhou2011}
\bibinfo{author}{Mou, D.} \emph{et~al.}
\newblock \bibinfo{title}{Distinct fermi surface topology and nodeless
  superconducting gap in a \textrm{(Tl$_{0.58}$Rb$_{0.42}$)Fe$_{1.72}$Se$_{2}$}
  superconductor}.
\newblock \emph{\bibinfo{journal}{Phys. Rev. Lett.}}
  \textbf{\bibinfo{volume}{106}}, \bibinfo{pages}{107001}
  (\bibinfo{year}{2011}).

\bibitem{JPHu2012}
\bibinfo{author}{Hu, J.~P.} \& \bibinfo{author}{Ding, H.}
\newblock \bibinfo{title}{Local antiferromagnetic exchange and collaborative
  fermi surface as key ingredients of high temperature superconductors}.
\newblock \emph{\bibinfo{journal}{Sci. Rep.}} \textbf{\bibinfo{volume}{2}},
  \bibinfo{pages}{381} (\bibinfo{year}{2012}).

\bibitem{XHChen43K}
\bibinfo{author}{Chen, X.~H.} \emph{et~al.}
\newblock \bibinfo{title}{Superconductivity at 43?{K} in
  {SmFeAsO$_{1-x}$F$_x$}}.
\newblock \emph{\bibinfo{journal}{Nature}} \textbf{\bibinfo{volume}{453}},
  \bibinfo{pages}{761--762} (\bibinfo{year}{2008}).

\bibitem{RenZA2008}
\bibinfo{author}{Ren, Z.-A.} \emph{et~al.}
\newblock \bibinfo{title}{Superconductivity at 55 {K} in iron-based {F}-doped
  layered quaternary compound {Sm[O$_{1-x}$F$_x$]FeAs}}.
\newblock \emph{\bibinfo{journal}{Chin. Phys. Lett.}}
  \textbf{\bibinfo{volume}{25}}, \bibinfo{pages}{2215} (\bibinfo{year}{2008}).

\bibitem{WANG2008538}
\bibinfo{author}{Wang, X.} \emph{et~al.}
\newblock \bibinfo{title}{The superconductivity at 18 {K} in {LiFeAs} system}.
\newblock \emph{\bibinfo{journal}{Solid State Commun.}}
  \textbf{\bibinfo{volume}{148}}, \bibinfo{pages}{538--540}
  (\bibinfo{year}{2008}).

\bibitem{PRB060505}
\bibinfo{author}{Tapp, J.~H.} \emph{et~al.}
\newblock \bibinfo{title}{{LiFeAs}: An intrinsic feas-based superconductor with
  ${T}_{c}=18\text{ }\text{K}$}.
\newblock \emph{\bibinfo{journal}{Phys. Rev. B}} \textbf{\bibinfo{volume}{78}},
  \bibinfo{pages}{060505} (\bibinfo{year}{2008}).

\bibitem{FeSe}
\bibinfo{author}{Hsu, F.-C.} \emph{et~al.}
\newblock \bibinfo{title}{Superconductivity in the {PbO}-type structure
  $\alpha$-{FeSe}}.
\newblock \emph{\bibinfo{journal}{Natl. Acad. Sci.}}
  \textbf{\bibinfo{volume}{105}}, \bibinfo{pages}{14262--14264}
  (\bibinfo{year}{2008}).

\bibitem{122}
\bibinfo{author}{Rotter, M.}, \bibinfo{author}{Tegel, M.} \&
  \bibinfo{author}{Johrendt, D.}
\newblock \bibinfo{title}{Superconductivity at 38 {K} in the iron arsenide
  {Ba$_{1-x}$K$_x$Fe$_2$As$_2$}}.
\newblock \emph{\bibinfo{journal}{Phys. Rev. Lett.}}
  \textbf{\bibinfo{volume}{101}}, \bibinfo{pages}{107006}
  (\bibinfo{year}{2008}).

\bibitem{Qi2008}
\bibinfo{author}{Qi, Y.} \emph{et~al.}
\newblock \bibinfo{title}{Superconductivity at 34.7?{K} in the iron arsenide
  {Eu$_{0.7}$Na$_{0.3}$Fe$_2$As$_2$}}.
\newblock \emph{\bibinfo{journal}{New J. Phys.}} \textbf{\bibinfo{volume}{10}},
  \bibinfo{pages}{123003} (\bibinfo{year}{2008}).

\bibitem{Zhu2009}
\bibinfo{author}{Zhu, X.} \emph{et~al.}
\newblock \bibinfo{title}{Transition of stoichiometric {Sr$_2$VO$_3$FeAs} to a
  superconducting state at 37.2 {K}}.
\newblock \emph{\bibinfo{journal}{Phys. Rev. B}} \textbf{\bibinfo{volume}{79}},
  \bibinfo{pages}{220512(R)} (\bibinfo{year}{2009}).

\bibitem{12442-1}
\bibinfo{author}{Wang, Z.-C.} \emph{et~al.}
\newblock \bibinfo{title}{Superconductivity in {KCa$_2$Fe$_4$As$_4$F$_2$} with
  separate double {Fe$_2$As$_2$} layers}.
\newblock \emph{\bibinfo{journal}{J. Am. Chem. Soc.}}
  \textbf{\bibinfo{volume}{138}}, \bibinfo{pages}{7856--7859}
  (\bibinfo{year}{2016}).

\bibitem{Wen_2008}
\bibinfo{author}{Wen, H.~H.}, \bibinfo{author}{Mu, G.}, \bibinfo{author}{Fang,
  L.}, \bibinfo{author}{Yang, H.} \& \bibinfo{author}{Zhu, X.}
\newblock \bibinfo{title}{Superconductivity at 25?{K} in hole-doped
  {(La$_{1-x}$Sr$_x$)OFeAs}}.
\newblock \emph{\bibinfo{journal}{Europhys. Lett.}}
  \textbf{\bibinfo{volume}{82}}, \bibinfo{pages}{17009} (\bibinfo{year}{2008}).

\bibitem{B815830D}
\bibinfo{author}{Kasperkiewicz, K.}, \bibinfo{author}{Bos, J.-W.~G.},
  \bibinfo{author}{Fitch, A.~N.}, \bibinfo{author}{Prassides, K.} \&
  \bibinfo{author}{Margadonna, S.}
\newblock \bibinfo{title}{Structural and electronic response upon hole doping
  of rare-earth iron oxyarsenides {Nd$_{1-x}$Sr$_x$FeAsO} (0$<x \leq$0.2)}.
\newblock \emph{\bibinfo{journal}{Chem. Commun.}} \bibinfo{pages}{707--709}
  (\bibinfo{year}{2009}).

\bibitem{Mu2009}
\bibinfo{author}{Mu, G.} \emph{et~al.}
\newblock \bibinfo{title}{Synthesis, structural, and transport properties of
  the hole-doped superconductor {Pr$_{1-x}$Sr$_x$FeAsO}}.
\newblock \emph{\bibinfo{journal}{Phys. Rev. B}} \textbf{\bibinfo{volume}{79}},
  \bibinfo{pages}{104501} (\bibinfo{year}{2009}).

\bibitem{CaFeAsF}
\bibinfo{author}{Matsuishi, S.} \emph{et~al.}
\newblock \bibinfo{title}{Superconductivity induced by {Co}-doping in
  quaternary fluoroarsenide {CaFeAsF}}.
\newblock \emph{\bibinfo{journal}{J. Am. Chem. Soc.}}
  \textbf{\bibinfo{volume}{130}}, \bibinfo{pages}{14428--14429}
  (\bibinfo{year}{2008}).

\bibitem{SrFeAsF-1}
\bibinfo{author}{Han, F.}, \bibinfo{author}{Zhu, X.}, \bibinfo{author}{Mu, G.},
  \bibinfo{author}{Cheng, P.} \& \bibinfo{author}{Wen, H.-H.}
\newblock \bibinfo{title}{{SrFeAsF} as a parent compound for iron pnictide
  superconductors}.
\newblock \emph{\bibinfo{journal}{Phys. Rev. B}} \textbf{\bibinfo{volume}{78}},
  \bibinfo{pages}{180503} (\bibinfo{year}{2008}).

\bibitem{SrFeAsF-2}
\bibinfo{author}{Tegel, M.} \emph{et~al.}
\newblock \bibinfo{title}{Synthesis, crystal structure and spin-density-wave
  anomaly of the iron arsenide-fluoride {SrFeAsF}}.
\newblock \emph{\bibinfo{journal}{Europhys. Lett.}}
  \textbf{\bibinfo{volume}{84}}, \bibinfo{pages}{67007} (\bibinfo{year}{2008}).

\bibitem{ChengEPL2009}
\bibinfo{author}{Cheng, P.} \emph{et~al.}
\newblock \bibinfo{title}{High-{T$_c$} superconductivity induced by doping
  rare-earth elements into {CaFeAsF}}.
\newblock \emph{\bibinfo{journal}{Europhys. Lett.}}
  \textbf{\bibinfo{volume}{85}}, \bibinfo{pages}{67003} (\bibinfo{year}{2009}).

\bibitem{Matsuishi2009}
\bibinfo{author}{Matsuishi, S.} \emph{et~al.}
\newblock \bibinfo{title}{Effect of 3d transition metal doping on the
  superconductivity in quaternary fluoroarsenide {CaFeAsF}}.
\newblock \emph{\bibinfo{journal}{New J. Phys.}} \textbf{\bibinfo{volume}{11}},
  \bibinfo{pages}{025012} (\bibinfo{year}{2009}).

\bibitem{Pressure2010}
\bibinfo{author}{Okada, H.} \emph{et~al.}
\newblock \bibinfo{title}{Pressure dependence of the superconductor transition
  temperature of {Ca(Fe$_{1-x}$Co$_x$)AsF} compounds: A comparison with the
  effect of pressure on {LaFeAsO$_{1-x}$F$_x$}}.
\newblock \emph{\bibinfo{journal}{Phys. Rev. B}} \textbf{\bibinfo{volume}{81}},
  \bibinfo{pages}{054507} (\bibinfo{year}{2010}).

\bibitem{Gao2018}
\bibinfo{author}{Gao, B.}, \bibinfo{author}{Ma, Y.}, \bibinfo{author}{Mu, G.}
  \& \bibinfo{author}{Xiao, H.}
\newblock \bibinfo{title}{Pressure-induced superconductivity in parent
  \textrm{CaFeAsF} single crystals}.
\newblock \emph{\bibinfo{journal}{Phys. Rev. B}} \textbf{\bibinfo{volume}{97}},
  \bibinfo{pages}{174505} (\bibinfo{year}{2018}).

\bibitem{Ma2015}
\bibinfo{author}{Ma, Y.~H.} \emph{et~al.}
\newblock \bibinfo{title}{Growth and characterization of millimeter-sized
  single crystals of \textrm{CaFeAsF}}.
\newblock \emph{\bibinfo{journal}{Supercond. Sci. Technol.}}
  \textbf{\bibinfo{volume}{28}}, \bibinfo{pages}{085008}
  (\bibinfo{year}{2015}).

\bibitem{Ma2016}
\bibinfo{author}{Ma, Y.~H.} \emph{et~al.}
\newblock \bibinfo{title}{Growth and characterization of
  \textrm{CaFe$_{1-x}$Co$_x$AsF} single crystals by {CaAs} flux method}.
\newblock \emph{\bibinfo{journal}{J. Cryst. Growth}}
  \textbf{\bibinfo{volume}{451}}, \bibinfo{pages}{161} (\bibinfo{year}{2016}).

\bibitem{Taichi2018}
\bibinfo{author}{Terashima, T.} \emph{et~al.}
\newblock \bibinfo{title}{Fermi surface with {D}irac fermions in
  \textrm{CaFeAsF} determined via quantum oscillation measurements}.
\newblock \emph{\bibinfo{journal}{Phys. Rev. X}} \textbf{\bibinfo{volume}{8}},
  \bibinfo{pages}{011014} (\bibinfo{year}{2018}).

\bibitem{Xiao2016}
\bibinfo{author}{Xiao, H.} \emph{et~al.}
\newblock \bibinfo{title}{Superconducting fluctuation effect in
  \textrm{CaFe$_{0.88}$Co$_{0.12}$AsF}}.
\newblock \emph{\bibinfo{journal}{J. Phys.: Condens. Matter}}
  \textbf{\bibinfo{volume}{28}}, \bibinfo{pages}{455701}
  (\bibinfo{year}{2016}).

\bibitem{CaFeAsFCo-Hc2}
\bibinfo{author}{Ma, Y.~H.} \emph{et~al.}
\newblock \bibinfo{title}{Strong anisotropy effect in iron-based superconductor
  \textrm{CaFe$_{0.882}$Co$_{0.118}$AsF}}.
\newblock \emph{\bibinfo{journal}{Supercond. Sci. Technol.}}
  \textbf{\bibinfo{volume}{30}}, \bibinfo{pages}{074003}
  (\bibinfo{year}{2017}).

\bibitem{Xu2018}
\bibinfo{author}{Xu, B.} \emph{et~al.}
\newblock \bibinfo{title}{Optical study of {D}irac fermions and related phonon
  anomalies in the antiferromagnetic compound \textrm{CaFeAsF}}.
\newblock \emph{\bibinfo{journal}{Phys. Rev. B}} \textbf{\bibinfo{volume}{97}},
  \bibinfo{pages}{195110} (\bibinfo{year}{2018}).

\bibitem{Ma2018}
\bibinfo{author}{Ma, Y.~H.} \emph{et~al.}
\newblock \bibinfo{title}{Magnetic-field-induced metal-insulator quantum phase
  transition in \textrm{CaFeAsF} near the quantum limit}.
\newblock \emph{\bibinfo{journal}{Sci. China Phys. Mech.}}
  \textbf{\bibinfo{volume}{61}}, \bibinfo{pages}{127408}
  (\bibinfo{year}{2018}).

\bibitem{Mu2018}
\bibinfo{author}{Mu, G.} \& \bibinfo{author}{Ma, Y.}
\newblock \bibinfo{title}{Single crystal growth and physical property study of
  1111-type {Fe}-based superconducting system \textrm{CaFeAsF}}.
\newblock \emph{\bibinfo{journal}{Acta Phys. Sin.}}
  \textbf{\bibinfo{volume}{67}}, \bibinfo{pages}{177401}
  (\bibinfo{year}{2018}).

\bibitem{Terashima2022-1}
\bibinfo{author}{Terashima, T.}, \bibinfo{author}{Uji, S.},
  \bibinfo{author}{Wang, T.} \& \bibinfo{author}{Mu, G.}
\newblock \bibinfo{title}{Topological frequency shift of quantum oscillation in
  {CaFeAsF}}.
\newblock \emph{\bibinfo{journal}{npj Quantum Mater.}}
  \textbf{\bibinfo{volume}{7}}, \bibinfo{pages}{25} (\bibinfo{year}{2022}).

\bibitem{Terashima2022-2}
\bibinfo{author}{Terashima, T.} \emph{et~al.}
\newblock \bibinfo{title}{Anomalous high-field magnetotransport in {CaFeAsF}
  due to the quantum hall effect}.
\newblock \emph{\bibinfo{journal}{npj Quantum Mater.}}
  \textbf{\bibinfo{volume}{7}}, \bibinfo{pages}{62} (\bibinfo{year}{2022}).

\bibitem{Shlyk2014}
\bibinfo{author}{Shlyk, L.} \emph{et~al.}
\newblock \bibinfo{title}{Crystal structure and superconducting properties of
  hole-doped {Ca$_{0.89}$Na$_{0.11}$FFeAs} single crystals}.
\newblock \emph{\bibinfo{journal}{Supercond. Sci. Technol.}}
  \textbf{\bibinfo{volume}{27}}, \bibinfo{pages}{044011}
  (\bibinfo{year}{2014}).

\bibitem{12442}
\bibinfo{author}{Wang, T.} \emph{et~al.}
\newblock \bibinfo{title}{Strong pauli paramagnetic effect in the upper
  critical field of {KCa$_2$Fe$_4$As$_4$F$_2$}}.
\newblock \emph{\bibinfo{journal}{Sci. China Phys. Mech. Astron.}}
  \textbf{\bibinfo{volume}{63}}, \bibinfo{pages}{227412}
  (\bibinfo{year}{2020}).

\bibitem{Werthamer1966}
\bibinfo{author}{Werthamer, N.~R.}, \bibinfo{author}{Helfand, E.} \&
  \bibinfo{author}{Hohenberg, P.~C.}
\newblock \bibinfo{title}{Temperature and purity dependence of the
  superconducting critical field, ${H}_{c2}$. iii. electron spin and spin-orbit
  effects}.
\newblock \emph{\bibinfo{journal}{Phys. Rev.}} \textbf{\bibinfo{volume}{147}},
  \bibinfo{pages}{295--302} (\bibinfo{year}{1966}).

\bibitem{WANG20231354223}
\bibinfo{author}{Wang, X.} \emph{et~al.}
\newblock \bibinfo{title}{Investigation of the pauli paramagnetic effect in
  systematically tuned {NbN} thin films}.
\newblock \emph{\bibinfo{journal}{Physica C}} \textbf{\bibinfo{volume}{606}},
  \bibinfo{pages}{1354223} (\bibinfo{year}{2023}).

\end{thebibliography}

%
%
%

\end{document}